# Low-Frequency Electronic Noise in Superlattice and Random-Packed Thin Films of Colloidal Quantum Dots

Adane Geremew[1], Caroline Qian[2], Alex Abelson[3], Sergey Rumyantsev[1,4], Fariborz Kargar[1], Matt Law[2,3,5] and Alexander A. Balandin[1,6]*

[1]Nano-Device Laboratory (NDL), Department of Electrical and Computer Engineering, University of California, Riverside, California 92521 USA

[2]Department of Chemical and Biomolecular Engineering, University of California, Irvine, California 92627 USA

[3]Department of Materials Science and Engineering, University of California, Irvine, California 92697 USA

[4]CENTERA Laboratories, Institute of High-Pressure Physics, Polish Academy of Sciences, Warsaw 01-142 Poland

[5]Department of Chemistry, University of California, Irvine, California 92697 USA

[6]Phonon Optimized Engineered Materials (POEM) Center, Materials Science and Engineering Program, University of California, Riverside, California 92521 USA

* Corresponding author (A.A.B.): balandin@ece.ucr.edu ; Web-site: https://balandingroup.ucr.edu/

## Abstract

We report measurements of low-frequency electronic noise in ordered superlattice, weakly-ordered and random-packed thin films of 6.5 nm PbSe quantum dots prepared using several different ligand chemistries. For all samples, the normalized noise spectral density of the dark current revealed a Lorentzian component, reminiscent of the generation-recombination noise, superimposed on the $1/f$ background ($f$ is the frequency). An activation energy of ~0.3 eV was extracted from the temperature dependence of the noise spectra. The noise level in the ordered films was lower than that in the weakly-ordered and random-packed films. A large variation in the magnitude of the noise spectral density was also observed in samples with different ligand treatments. The obtained results are important for application of colloidal quantum dot films in photodetectors.

Solution-processed quantum dot (QD) optoelectronic devices may offer low cost, large area, mechanically flexible and manufacturable large-scale device integration.[1–5] Solution-based processes include spin coating, dip coating, Langmuir-Schaeffer deposition, spraying and inkjet printing. Typically, the performance of solution-processed devices is inferior to the performance of devices fabricated by conventional techniques. However, the low cost, scalability and other benefits make solution-processed optoelectronics attractive for a range of applications, including photodetectors, light emitting diodes and solar cells.[5–12] Colloidal QDs can be used to prepare random-packed or ordered QD thin films. Spatially-ordered QD assemblies are often called quantum dot superlattices (QD SLs).[1] The optical and electronic properties of QD SLs depend not only on the intrinsic characteristics of QDs but also on the QD packing density, orientation, inter-QD distance and dielectric medium. Tunable electronic band structures make QD SLs attractive for detector and photovoltaic applications.[5,13,14] The low thermal conductivity of QD films also suggests applications in thermoelectrics.[1,15]

It is predicted theoretically that QD SLs with small QD size and inter-dot distance and low levels of defects and disorder offer attractive possibilities for controlling the electronic band structure and acoustic phonon dispersion.[14,16] Strong electron wave function overlap in QD SLs can lead to formation of electronic mini-bands, and, as a result, substantially higher charge carrier mobility than is achievable in films of otherwise-comparable random-packed QDs. The long-range order of QDs is essential for formation of mini-bands and emergence of band transport instead of the hopping transport characteristic of random QD films. Long-range order can also lead to strong modification of the acoustic phonon dispersion, with corresponding changes in electron-phonon scattering and light-matter interactions.[12,14–17] For more than two decades, the efforts in synthesis and testing of QD SLs synthesized by molecular beam epitaxy,[18–20] solution processing[21–23] and other techniques[24] were focused on improving the long-range order to achieve formation of coherent mini-bands and, correspondingly, enhanced electron mobility and modified optical response.[1] There have been only a few studies of current fluctuation and noise processes in QD films and devices.[25–28] We are aware of only one detailed report on low-frequency noise in colloidal QD films.[27] Knowledge of the low-frequency noise characteristics of QD films is important from both the fundamental and applied points of view. Noise characteristics can provide insight into charge transport and tunneling mechanisms in QD films. Understanding of noise mechanisms and development of noise reduction approaches are important for practical applications of QD films in photodetectors.

In this Letter, we report low-frequency noise measurements in three types of QD thin films: highly-ordered QD SLs (SL films), weakly-ordered spin-cast films (SC films), and random-packed dip-coated films (DC films). We find that the SL films have less noise than the SC and DC films. The difference in the noise spectral density between the SL and DC films varies from a factor of two to more than two orders of magnitude at room temperature (RT). The noise levels of DC films with different ligand chemistries, *i.e.*, films prepared with ethylenediamine versus ammonium thiocyanate, span more than an order of magnitude. One important finding is that the spectra of all films show a Lorentzian component superimposed on the 1/$f$ background, reminiscent of generation-recombination (G-R) noise. Interestingly, a single activation energy of ~0.3 eV was extracted from the noise temperature dependence of SL and DC films, prepared by different chemistries. The obtained results have important implications for proposed applications of QDs in photodetectors because low-frequency noise often limits the detectivity and selectivity of photodetectors and sensors.[29–34] Our observation of lower noise in QD SLs provides additional motivation for research to improve the long-range order of colloidal QD superlattices.

In this study, we fabricated 30-70 nm thick films of 6.5 nm PbSe QDs using three different methods and ligand chemistries in order to study the impact of spatial order and surface chemistry on low-frequency noise (see Table I). All films were infilled and overcoated with a 20 nm thick layer of amorphous aluminum oxide via atomic layer deposition (ALD) to prevent oxidation of the QDs.[35] To investigate the role of spatial order, we fabricated epitaxial superlattice (epi-SL) films with ~250 nm lateral SL grain sizes via self-assembly of QDs on a liquid surface.[36,37] These films contain a mixture of adsorbed ethylene glycoxide, iodide and residual oleate surface ligands. The second type of sample (the SC films) feature ~25 nm lateral superlattice grains and similar surface ligands and coverage as the SL films. These films have very similar surface chemistry to the epi-SL films. On a sub-100 nm length scale, the SL films possess more uniform inter-QD distances and connectivity due to oriented attachment (*i.e.*, epitaxial fusion of the QDs) in three dimensions.[37,38] At a length scale below 10 μm, the SC films are smooth and continuous whereas the SL films have more significant macroscopic cracking that occurs during QD self-assembly. Scanning electron microscope (SEM) images highlighting these differences are provided in Figure 1. The third type of sample was random-packed QD films prepared by dip coating (DC films) using either ethylenediamine (DC EDA) or ammonium thiocyanate (DC SCN #1 and #2) ligand treatments. These films were deposited using a layer-by-layer dip coating process that yields optically smooth, continuous, and dense films (see Figure 1).[39] The DC SCN films contain adsorbed thiocyanate and the DC EDA films possess a mixture of oleate and ethylenediamine ligands. Fourier transform infrared (FTIR) spectra shown in the Supporting Information highlight the differences in surface chemistry between the samples studied here.

[Figure 1]

For electrical and noise measurements, we prepared QD films on SiO$_2$/Si wafers pre-patterned with metal contacts. Ti/Au contacts (5 nm / 35 nm) were separated by a channel with a length of 25 μm and width of 1000 μm, as defined by conventional photolithography The quantum-confined band gap of the QDs in solution was 0.69 eV, as expected for 6.5 nm PbSe QDs. The bulk band gap of PbSe is 0.29 eV at RT.[40,41] All of the devices showed *n*-channel behavior after ALD infilling, with an electron mobility in the range of 1-4 cm$^2$ V$^{-1}$ s$^{-1}$, which is typical for such materials.[1,3] Figure 2 (a) shows a schematic of the fabricated devices. The current-voltage (*I-V*) characteristics of the devices were measured with a semiconductor parameter analyzer (Agilent B1500). Figure 2 (b) shows the electrical resistivity, $\rho$, for representative QD films of all three types (SL, SC, and DC) as a function of temperature, *T*. The sample names in the legend correspond to those in the Table I. Since the focus of the present study is on the dark current noise characteristics, the data in Figure 2 (b) were measured without illumination.

[Figure 2]

The resistivity of the SL films was smaller than that of the SC and DC films (see Figure 2 (b)). For all three types of samples, the resistivity decreases with temperature, suggesting that transport occurs by phonon-assisted tunneling, *i.e.*, hopping.[42–45] The hopping conductance is commonly analyzed using the expression[43]

$$G = G_o e^{-(T_o/T)^p},$$

where $T_0$ depends on the material properties and the localization length in the given structure and $p$ is a parameter defined by the type of hopping. The resistivity data analysis using Arrhenius plots and the $\ln(G)$ *vs.* $\ln(T)$ dependence (see Supplemental Materials) indicate nearest-neighbor hopping (NNH) in the high-temperature region and variable-range hopping (VRH) at lower temperature. Similar temperature dependences of the conductivity for PbSe QDs of the same diameter were reported in Ref.[37]. From the Arrhenius plot, we extracted activation energies of 0.171 eV and 0.137 eV for NNH transport in the DC and SL films, respectively.

The noise spectra were determined with a dynamic signal analyzer (Stanford Research) with inbuilt low-noise amplifier. The devices were DC biased with a "quiet" battery-potentiometer circuit in order to minimize 60 Hz noise from the electrical grid. The noise measurements were conducted in a two-terminal device configuration. Details of our noise measurement procedures have been reported elsewhere.[46–48] In Figure 3 (a), we present the normalized current noise spectral density, $S_I/I^2$, as a function of frequency, $f$, at different temperatures ($I$ is the current through the device) for a representative SL and DC QD film. The measurements were conducted at a source-drain bias of 1.0 V. Figure 3 (b) shows $S_I/I^2$ as a function of frequency at different bias voltages. In the studied set of samples, the QD SL produces less noise than the DC QD film at all bias voltages and temperatures. For some $f$ and $T$, the difference in the noise level, $S_I/I^2$, is more than an order of magnitude.

[Figure 3]

For all samples, we examined the noise spectral density scaling with the current (see Figure 4 (a) and (b)). The noise in all samples followed the $S_I \sim I^2$ trend with only small deviations. This indicates that the electrical current does not induce strong Joule heating, annealing or other structural or morphological changes.[49] This is in contrast to a previous report of low-frequency noise in colloidal QD films, which revealed strong deviation from the $S_I \sim I^2$ dependence.[27] In Figure 4 (c), we present the normalized noise spectral density, $S_I/I^2$, as a function of temperature. The noise level in the SL films is the lowest of all examined samples. However, at certain temperature and bias ranges, the noise spectral density in SL films becomes rather close to that in the random DC films. The difference in the noise level between DC and SC films is also large. For this reason, it is difficult to establish from these data the relative importance of spatial order, ligand chemistry, and other factors determining the noise level in QD films.

[Figure 4]

As one can see from Figure 3 (a) and Figure 4 (c), both the amplitude of noise and shape of the spectra depend on temperature. Similar to G-R noise in semiconductors, which appears as Lorentzian peaks, this can be a result of a random process with a well-defined characteristic time that depends on temperature.[29,50–52] In general, the spectral density of G-R noise is described by the Lorentzian: $S_I(f)=S_0/[1+(2\pi f\tau)^2]$, where $S_0$ is the frequency independent portion of $S_I(f)$ observed at $f<<f_c=(2\pi\tau)^{-1}$ and $\tau$ is the time constant associated with the return to equilibrium of the occupancy of the trap level. In typical semiconductors, the spectral density of the G-R noise is often expressed as,[50]

$$\frac{S_I}{I^2} = \frac{4N_t}{Vn^2}\frac{\tau F(1-F)}{1+(\omega\tau)^2},\tag{1}$$

where $\omega = 2\pi f$, $V$ is the sample volume, $n$ is the equilibrium electron concentration for an $n$-type material, and $F$ is the trap state occupancy function. The G-R noise time constant, $\tau$, can be further related to the trap state capture and release time constants. The most common description of $1/f$ noise, dominated by fluctuations in the number of charge carriers, $N$, stems from the observation that a superposition of individual G–R noise sources with the lifetime distributed on a exponentially wide timescale, within the $\tau_1$ and $\tau_2$ limits, gives the $1/f$ type spectrum in the intermediate range of frequencies $1/\tau_2 < \omega < 1/\tau_1$. If one specific G-R noise source, e.g. trap with the well-defined $\tau$, dominates the noise spectrum owing to its much higher concentration, then the Lorentzian associated with this trap appears superimposed over the 1/f background. Our experimental observation appears to be in line with this mechanism. In order to characterize this kind of process, it is common to plot the normalized noise spectral noise density multiplied by frequency, $S_I/I^2 \times f$, versus frequency. The position of the maximum, $f_c$, of this dependence defines the characteristic time of the random process, $\tau = 1/2\pi f_c$, at a given temperature. If the characteristic time depends exponentially on temperature, an Arrhenius plot allows extraction of the activation energy of this random process. However, such a method is inapplicable if the position of the maximum is outside of the studied frequency range or not clearly observed due to the $1/f$ noise background.

From our data, one can see that at low temperatures and low frequencies, the shape of the measured spectra is close to $1/f^2$, indicating the presence of a Lorentzian component with characteristic frequency below 1 Hz, *i.e.*, below the limits of our experimental setup. At high temperatures, the Lorentzian components are barely noticeable because they are masked by the $1/f$ noise. For this reason, we used an alternative approach for finding the characteristic time, $\tau$, by plotting the noise spectral density as a function of temperature at different frequencies.[53,54] If these dependences have maxima, it is assumed that $\tau = 1/2\pi f_c$ at the temperature of the maximum, $T_m$. The Arrhenius plot of $\ln(f_c)$ versus $1/T_m$ allows one to find the activation energy for the noise process. Figure 5 (a-b) show $S_I/I^2$ at different frequencies as a function of temperature. The dependences in Figure 5 (a) and (b) have clear maxima shifting with temperature, reflecting the temperature dependence of $f_c$.

[Figure 5]

Figure 6 shows the plot of $\ln(f_c)$ vs. $1/T_m$ that was used to extract the noise activation energies. The activation energy for the SL and DC films is nearly equal at ~0.3 eV. This activation energy is significantly higher than the activation energy of the NNH conductivities. Therefore, the temperature dependences of the conductivity and noise appear to be regulated by different mechanisms. This is a rather common situation in semiconductors. For example, G-R noise with a strong temperature dependence is often observed even when the conductivity is temperature independent.[43,49,55–57]. The SL and DC samples have been fabricated by different methods and have different ligands. The same activation energy of the noise in these devices can be an indication of the same noise mechanism. The activation energy of 0.3 eV is close to the bandgap of bulk PbSe. One of the possibilities is that the G-R-type random process can be related to the exchange of charge carriers between the shallow donors and acceptors within the same or neighboring QDs. The shallow donor states are close to the conduction band while the shallow acceptor states are close to the valence band. The energy difference between them is close to the bulk band gap. From the other side, the energies close to 0.3 eV can also be associated with the certain ligands or chemistries used in QD treatment (see Supplemental

Materials). The variation in the noise level in the ordered and random QD films can also be associated with the number of conducting channels in the QD films, which depend on the inter-dot distance, presence of cracks and variation of the QD density. Under the realistic assumption of independent fluctuators uniformly distributed in the sample, the normalized noise spectral density is inversely proportional to the volume of the conducting channels, $S_I/I^2 \sim 1/V$.[29,49] This leads to higher noise in QD films, which have fewer conducting channels. However, more experimental studies are required to establish the exact mechanism of the noise in such samples and discriminate the effect of differences in inter-QD distances and coupling, QD stoichiometry, surface doping, ligand coverage, and grain boundaries in QD films. We note that no theoretical models for low-frequency noise in QD films or superlattices exist at the moment. The conventionally accepted noise models are either for electron band conduction in semiconductors[29,49] and metals[29,58,59] or electron hopping in disordered semiconductor systems.[43,60] Ordered and random-packed QD films are a unique class of materials that will require dedicated investigation. One should also note that the signal-to-noise ratio of a photodetector system limited by 1/*f* noise cannot be improved by extending the measuring time, $t \propto 1/f$. The total accumulated energy of the flicker 1/*f* noise increases at least as fast as *t*. This consideration adds a practical motivation to more detail studies of low-frequency noise in colloidal QDs.

In conclusion, we reported on measurements of the low-frequency electronic noise in ordered, weakly-ordered, and random-packed films of colloidal quantum dots. An important finding is that the normalized noise spectral density of the dark current contains a Lorentzian component superimposed on the 1/*f* background that is reminiscent of G-R noise. An activation energy of ~0.3 eV was extracted from the noise spectrum temperature dependence for both superlattice and random-packed quantum dot films. The noise level in the ordered films was lower than that in the weakly-ordered and random-packed films. However, the measurements also reveal a large variation in noise levels between random-packed films prepared with different ligand treatments. The obtained results are important for application of colloidal quantum dot films in photodetectors.

**Methods**

*Materials.* All chemicals were used as received unless otherwise noted. Lead oxide (PbO, 99.999%), lead iodide ($PbI_2$, 99.9985%), and selenium shot (99.999%) were purchased from Alfa Aesar. Oleic acid (OA, technical grade, 90%), diphenylphosphine (DPP, 98%), 1-octadecene (ODE, 90%), anhydrous ethylene glycol (EG, 99.8%), anhydrous acetonitrile (99.99%), anhydrous hexanes (99%), anhydrous toluene (99.8%), 3-mercaptopropyltrimethoxysilane (3-MPTMS, 95%), trimethylaluminum (TMA, 97%), ammonium thiocyanate ($NH_4SCN$, 99.99%), and anhydrous dimethyl sulfoxide (DMSO, 99.9%) were purchased from Sigma Aldrich. Anhydrous 1,2-ethylenediamine (EDA, >98.0%) was purchased from TCI. Trioctylphosphine (TOP, technical grade, >90%) was acquired from Fluka and mixed with selenium shot for 24 hours to form a 1 M TOP-Se stock solution. 18.2 MΩ water (Milli-Q Gradient) was used for substrate cleaning and atomic layer deposition (ALD). Water for ALD was degassed with three freeze-pump-thaw cycles before use.

*Quantum dot synthesis.* PbSe QDs were synthesized and purified using standard air-free techniques. PbO (1.50 g), OA (5.00 g), and ODE (10.00 g) were mixed and degassed in a three-neck round-bottom flask at room temperature. Then the mixture was heated at 110°C under vacuum to form $Pb(OA)_2$ and dry the solution. After 1.5 hours, the $Pb(OA)_2$ solution was heated to 180°C under argon flow and 9.5 mL of a 1 M solution of TOP-Se containing 200 $\mu$L of DPP was rapidly injected into this hot solution. An immediate darkening of the solution was observed, and the QDs were grown for 105 seconds at ~160°C. The reaction was quenched with a liquid nitrogen bath and injection of 10 mL of anhydrous hexanes. The QDs were purified in an $N_2$-filled glovebox (<0.5 ppm $O_2$) by adding 15 mL of acetonitrile to the reaction solution, collecting the QDs by centrifugation, performing six cycles of redispersion/precipitation using toluene/acetonitrile (3 mL/22 mL), and then drying and storing the QDs as a powder in the glovebox.

*Device fabrication.* Pre-patterned $Si/SiO_2$ substrates were cleaned by 10 minute rounds of sonication in acetone, Millipore water, and isopropanol, then blown dry. Following cleaning, they were immersed in a 100 mM solution of 3-MPTMS in toluene for several hours, then rinsed with neat toluene and blown dry. Dip-coated films (DC films) were prepared by 10 sequential rounds of dipping pre-patterned substrates in: 1) a 4 g/L PbSe QD solution in hexanes for 1 second; 2) ligand exchange solution for 10 seconds; 3) a neat acetonitrile rinse for 3 seconds. For SCN films, a 15 mM $NH_4SCN$ solution in acetonitrile was used. For EDA films, a 1 M EDA solution in acetonitrile was used. Spin-coated films (SC films) were prepared by spin-coating substrates with a 25 g/L PbSe QD solution in octane at a speed of 2000 rpm for 40 seconds. The films were then soaked in a fresh 105 mM EDA solution in ethylene glycol for ~1 minute, followed by a 5 minute soak in $PbI_2$ in DMSO, then rinsed clean in neat DMSO and acetonitrile. This process was repeated twice. SL films were prepared by self-assembly of 60 μL of 30 g/L PbSe QD solution suspended in hexanes on a liquid ethylene glycol (EG) surface in a Teflon well. After solution deposition, the well was covered and the hexanes allowed to evaporate over the course of ~25 min, leaving a dried QD SL film floating on the EG. The well was then uncovered and 100 μL of 7.5 M EDA in acetonitrile (105 mM EDA concentration overall in the well) was injected into the EG, underneath the edge of the film, to initiate ligand exchange. After ~30 seconds, the section of film nearest to the EDA injection point was manually stamp transferred to the pre-patterned $Si/SiO_2$ substrate using a vacuum wand. The stamped film was then rinsed in clean acetonitrile and blown dry with flowing $N_2$. Lastly, the film was soaked in a 10 mM $PbI_2$ in DMSO solution for 5 min before being rinsed in clean DMSO and acetonitrile, and again blown dry under flowing $N_2$. Atomic layer deposition was performed at 55 °C in homemade ALD system at a base pressure of 200 mTorr. ~20 nm thick films were produced after 200 cycles of alternating TMA and $H_2O$ pulses. Dose times of 20 ms and wait times of 45 s were used for each precursor.

**ACKNOWLEDGEMENTS**

The work of A.G., F.K. and A.A.B. at UCR was supported, in part, by DARPA project W911NF18-1-0041 on quantum dot phonon engineered materials. C.Q., A.A., and M.L. at UCI were supported by the UC Office of the President under the UC Laboratory Fees Research Program Collaborative Research and Training Award LFR-17-477148. Some materials characterization was performed at the user facilities of the UC Irvine Materials Research Institute (IMRI). The authors acknowledge Dr. Ece Aytan and Dr. Ruben Salgado for their help with materials characterization at the UC Riverside Center for Nanoscale Science and Engineering (CNSE). The work of S.R. was partially supported by the "International Research Agendas" program of the Foundation for Polish Science co-financed by the European Union under the European Regional Development Fund (MAB/2018/9).

**FIGURE CAPTIONS**

**Figure 1:** Scanning electron microscope images of QD films with varying degrees of long-range order. (a – b) Dip-coated films, corresponding to samples "DC SCN #1", "DC SCN #2" and "DC EDA #1" in Table I. (c – d) Spin-coated films with short-range order, corresponding to sample "SC EDA #1" in Table I. (e – f) Epi-superlattice films with long-range order, corresponding to samples "SL #1" and "SL #2" in Table I.

**Figure 2:** (a) Schematics of the QD devices showing perspective (top panel) and cross-sectional (bottom panel) views. (b) Electrical resistivity of the SL, SC, and DC films as a function of temperature. The decrease in resistivity with increasing temperature is consistent with hopping transport.

**Figure 3:** (a) Normalized current noise spectral density, $S_I/I^2$, as a function of frequency at different temperatures. The data are presented for two samples: ordered SL #1 (dashed lines) and random DC SCN #1 (solid lines). (b) Normalized current noise spectral density, $S_I/I^2$, as a function of frequency for different source-drain biases, shown for the same devices as in (a). The noise level in the ordered SL film is consistently lower than that in the random DC films.

**Figure 4**: (a-b) Noise spectral density, $S_I$, as a function of current, $I$. The data are presented in two panels to clearly show the difference between the SL and DC films that had relatively close noise levels. Note that the slope is proportional to $\sim I^2$ for all samples. (b) Normalized noise spectral density, $S_I/I^2$, as a function of temperature, $T$, for the examined QD films. The noise spectral density was measured at $f = 10$ Hz in both (a) and (b).

**Figure 5:** (a) Normalized current noise spectral density, $S_I/I^2$, as a function of temperature for different frequencies for the ordered sample SL #1. (b) The same as in (a) for the random sample DC SCN #1. The blue arrows are guides to the eye, indicating the shift in the maximum of the noise spectral density with temperature. The dotted lines illustrate the process of finding the noise maximum for each temperature.

**Figure 6:** Arrhenius plot, ln($f_c$) vs 1/$T$, for the samples SL #1 and DC SCN #1. The similarity of the activation energies extracted from the noise spectra suggest the same noise mechanism in different types of QD films.

**Table I: Characteristics of the colloidal quantum dot thin films**

| Samples Name | Film Type | Ligand Treatment | $S_I/I^2$ (Hz$^{-1}$) T =100 K f=10 Hz | $S_I/I^2$ (Hz$^{-1}$) T = 300 K f=10 Hz |
|---|---|---|---|---|
| DC SCN #1 | dip coated (DC); random-packed QDs | ammonium thiocyanate (SCN) | $5.80 \times 10^{-9}$ | $1.85 \times 10^{-10}$ |
| DC SCN #2 | dip coated (DC); random-packed QDs | ammonium thiocyanate (SCN) | $3.98 \times 10^{-6}$ | $4.38 \times 10^{-10}$ |
| DC EDA #1 | Dip coated (DC); random-packed QDs | ethylenediamine in acetonitrile | $2.31 \times 10^{-9}$ | $2.63 \times 10^{-8}$ |
| SC EDA #1 | spin coated (SC); weakly-ordered QDs | ethylenediamine in ethylene glycol+ PbI$_2$ in dimethylsulfoxide | $1.04 \times 10^{-9}$ | $1.42 \times 10^{-8}$ |
| SL #1 | epi-superlattice; long-range order | ethylenediamine in ethylene glycol+ PbI$_2$ in dimethylsulfoxide | $5.86 \times 10^{-11}$ | $2.91 \times 10^{-11}$ |
| SL #2 | epi-superlattice; long-range order | ethylenediamine in ethylene glycol+ PbI$_2$ in dimethylsulfoxide | $2.43 \times 10^{-10}$ | $6.32 \times 10^{-11}$ |

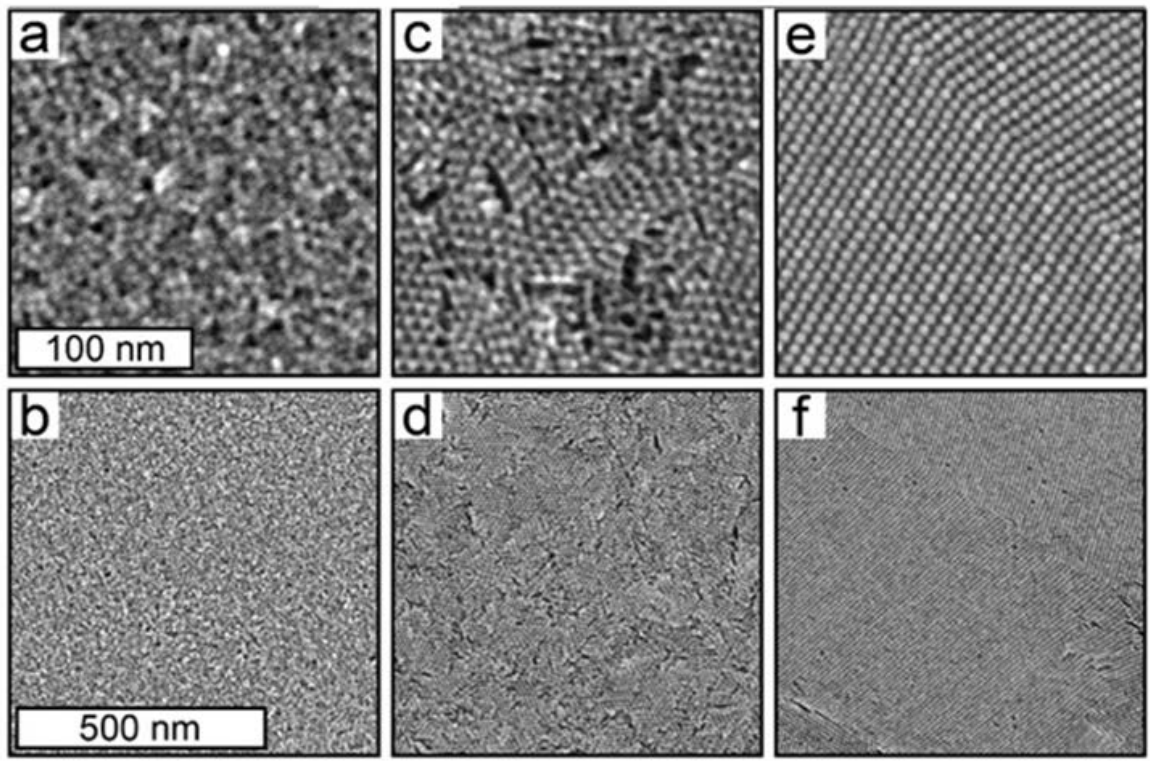

Figure 1

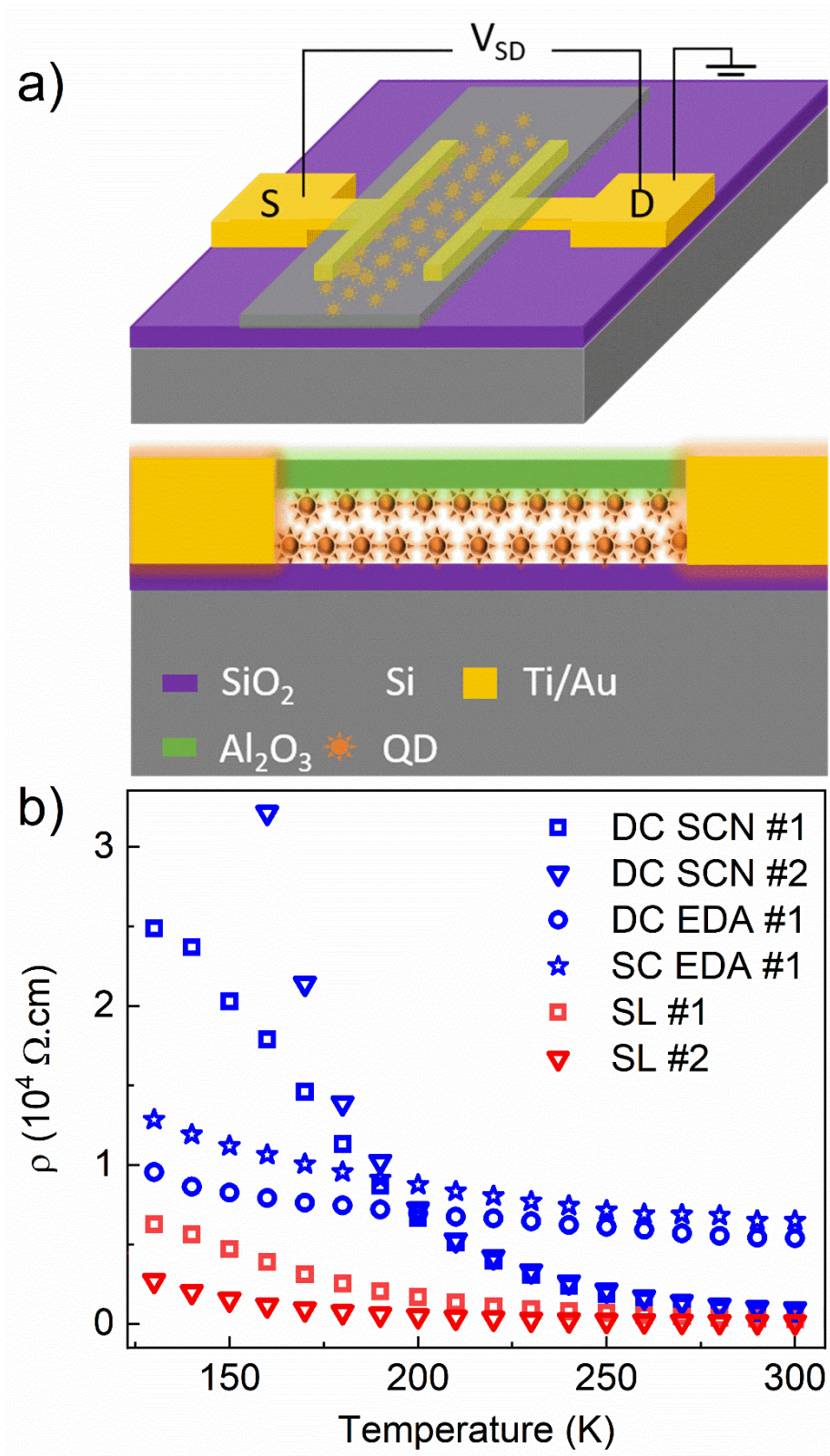

Figure 2

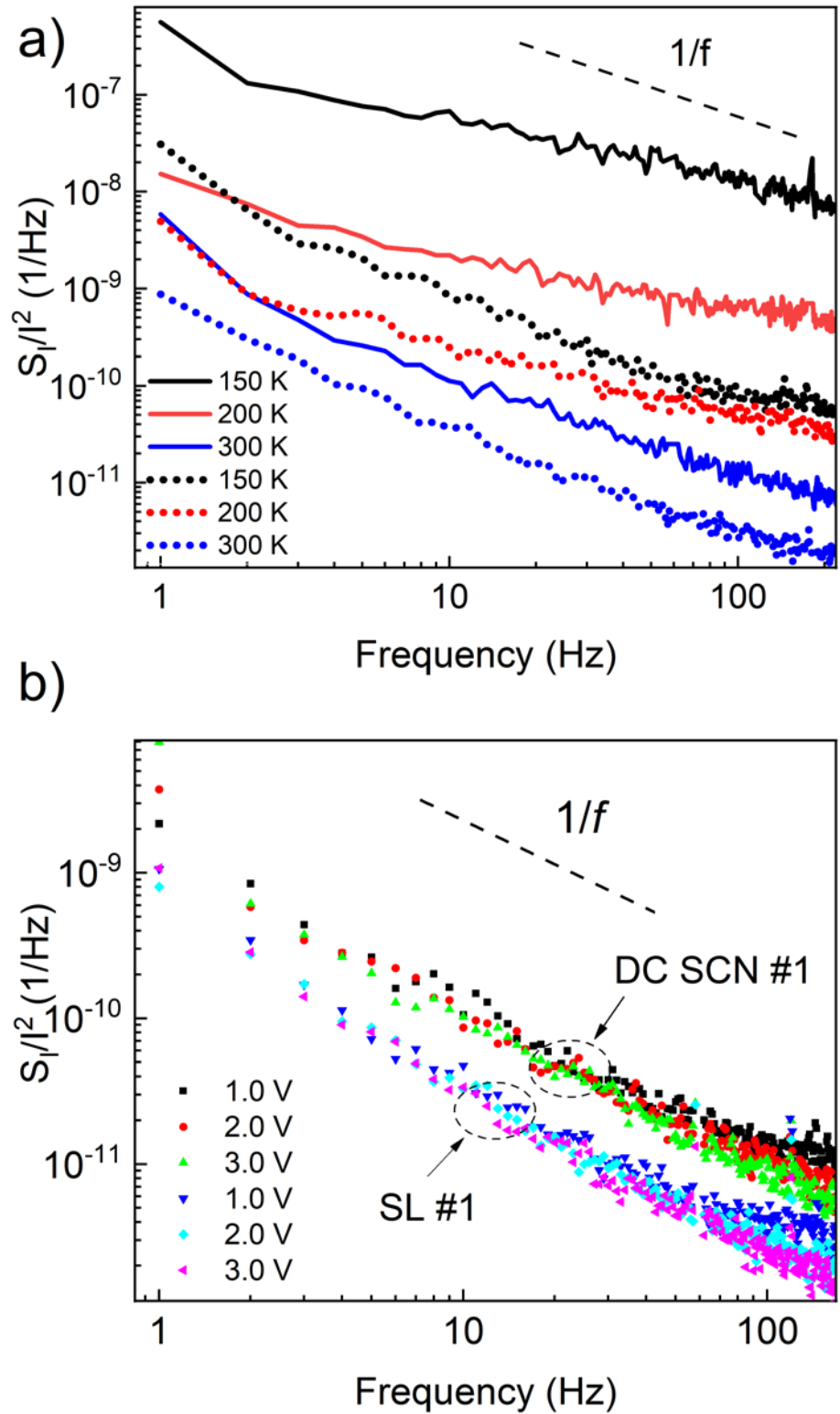

Figure 3

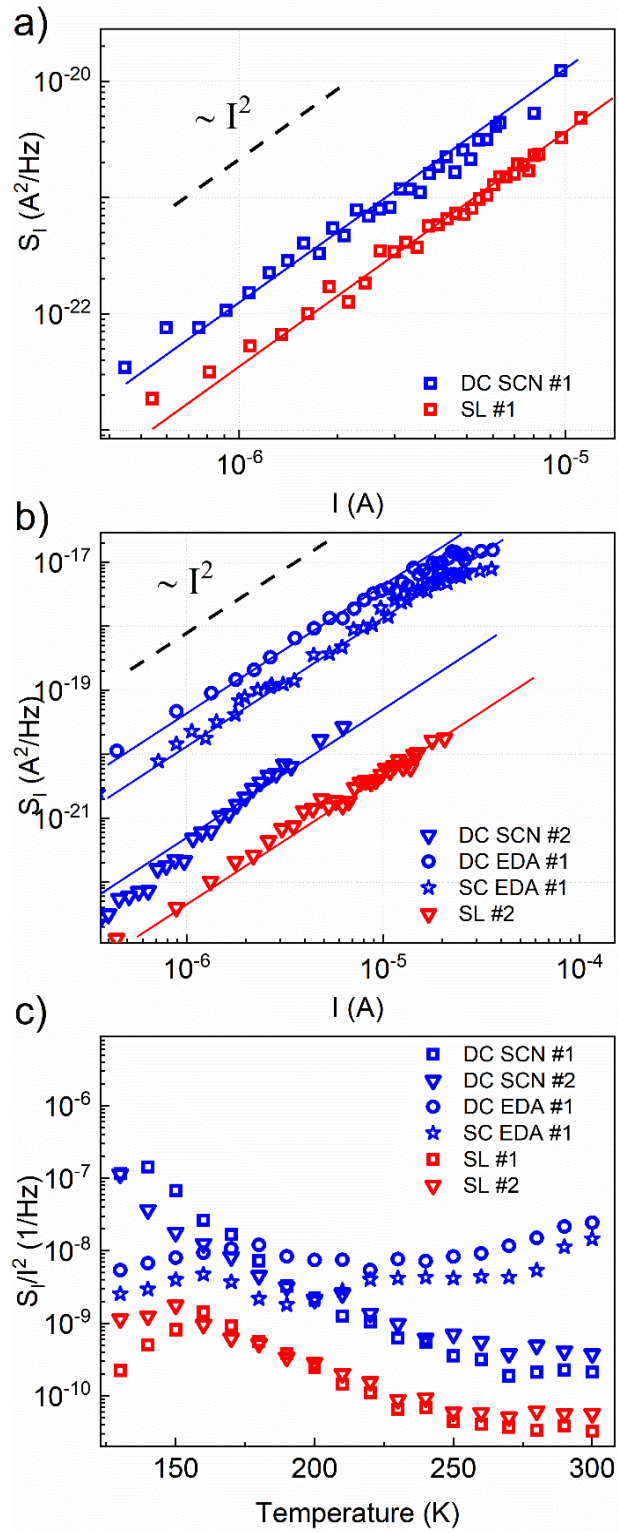

Figure 4

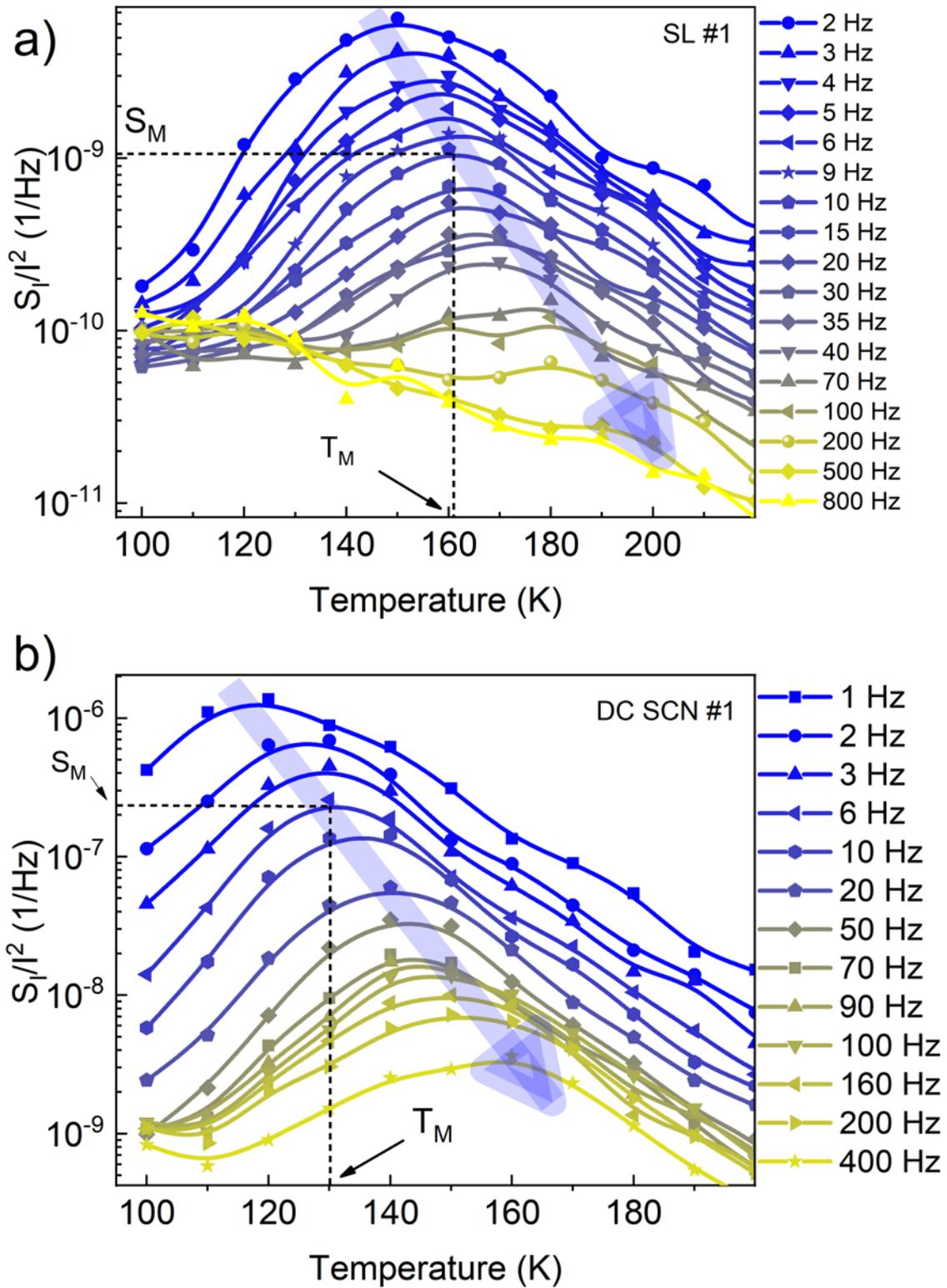

Figure 5

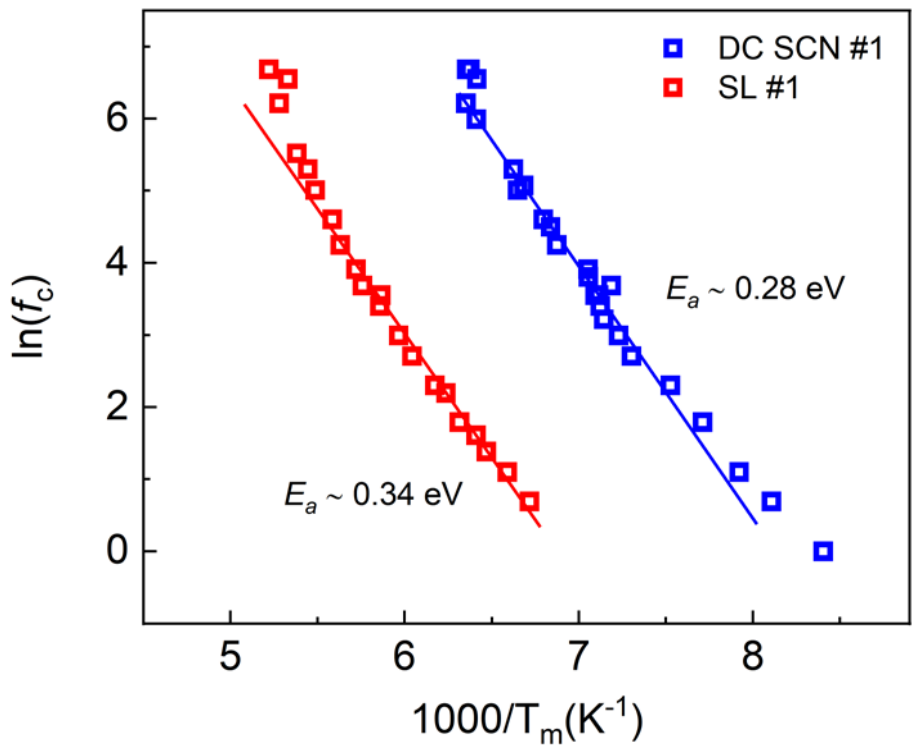

Figure 6

# Supplemental Information (SI)

## Low-Frequency Electronic Noise in Superlattice and Random-Packed Thin Films of Colloidal Quantum Dots

Adane Geremew[1], Caroline Yu Qian[2], Alex Abelson[2], Sergey Rumyantsev[3], Fariborz Kargar[1], Mathew Law[2] and Alexander A. Balandin[1*]

[1]Nano-Device Laboratory (NDL) and Phonon Optimized Engineered Materials (POEM) Center, Department of Electrical and Computer Engineering, University of California, Riverside, California 92521 USA

[2]Department of Chemistry, University of California, Irvine, California 92697 USA

[3]Center for Terahertz Research and Applications (CENTERA), Institute of High-Pressure Physics, Polish Academy of Sciences, Warsaw 01-142 Poland

[*] Corresponding author (A.A.B.): balandin@ece.ucr.edu ; Web-site: https://balandingroup.ucr.edu/

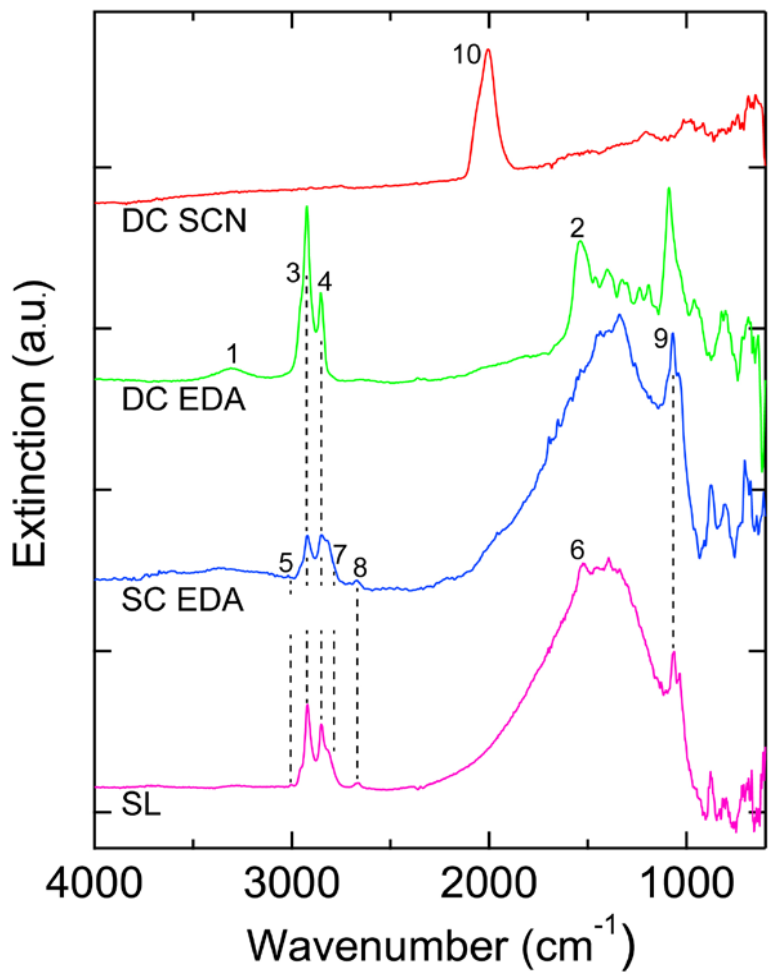

**Figure S1:** Fourier transform infrared spectra of the QD films in this study. Spectra of typical DC SCN, DC EDA, SC EDA, and SL films on silicon substrates. All samples were measured prior to ALD infilling. The labeled peaks were assigned as follows: peaks 1-2 are adsorbed ethylenediamine (1: $v(NH_2)$, 2: $NH_2$ scissor), peaks 3-4 are adsorbed oleate and ethylenediamine (3: $v_{as}(CH_2)$, 4: $v_s(CH_2)$), peaks 5-6 are unique to oleate (5: $v(HC=CH)$, 6: $v_s(COO^-)$), peaks 7-9 are adsorbed ethylene glycoxide (7: $v_{as}(CH_2)$, 8: $v_s(CH_2)$, 9: $v(C-O)$ and $v(C-C)$), and peak 10 comes from adsorbed thiocyanate (10: $v(C\equiv N)$). The ligand content of each film is summarized in Table S1.

**Table S1: Ligand content of the films**

| film type | oleate | glycoxide | ethylenediamine | thiocyanate |
| --- | --- | --- | --- | --- |
| DC SCN | no | no | no | yes |
| DC EDA | residual* | no | yes | no |
| SC EDA | yes | yes | no | no |
| SL | yes | yes | no | no |

*\* The presence of the ligand cannot be ruled out, but if present it is at a low concentration.*

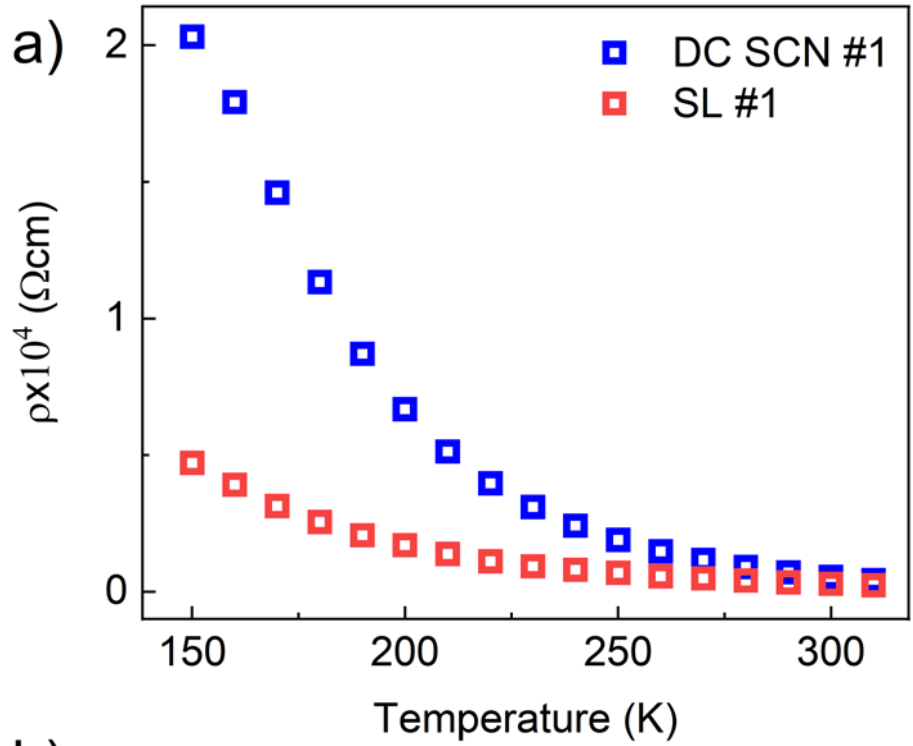

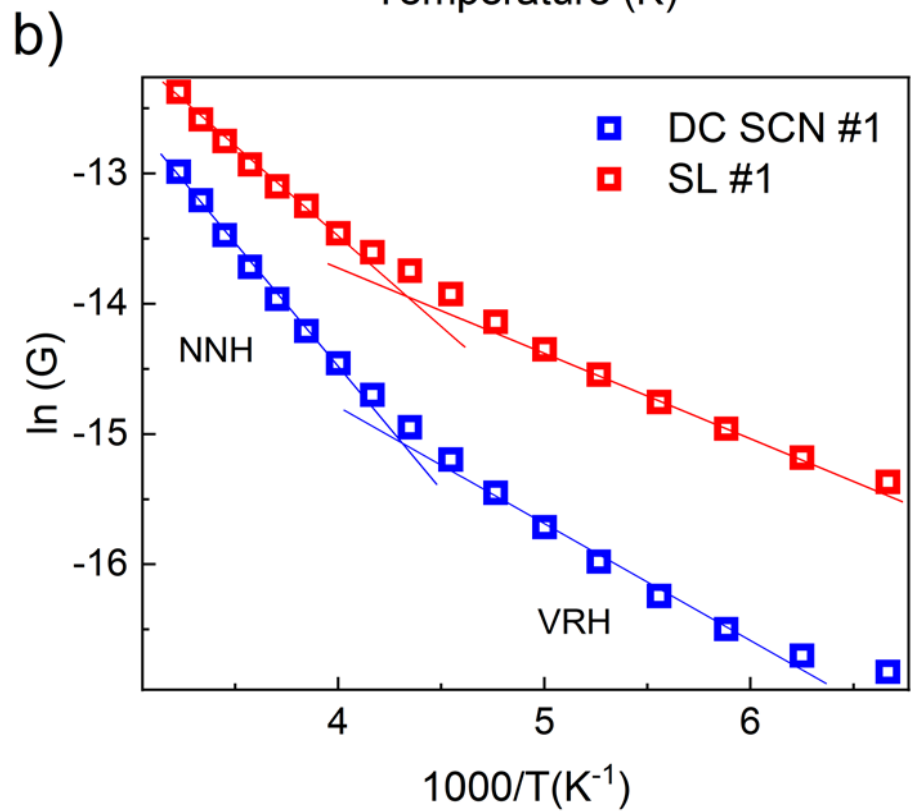

**Figure S2**: (a) Resistivity as a function of temperature. (b) Resistivity as a function of the inverse temperature.

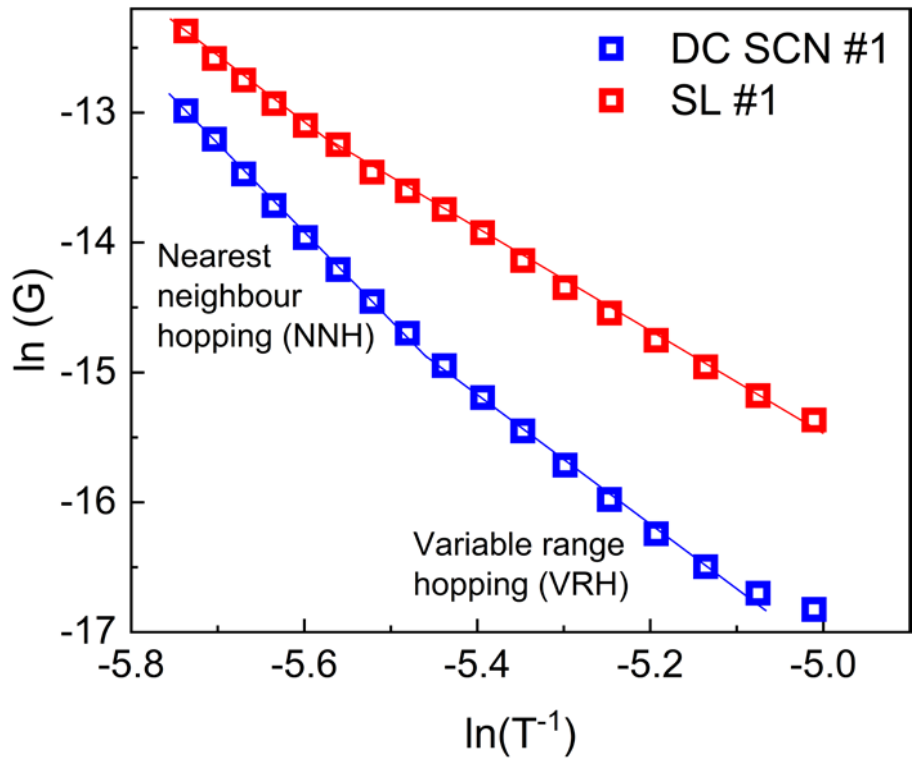

**Figure S3**: Logarithmic plot of the electrical conductivity as a function of inverse temperature.

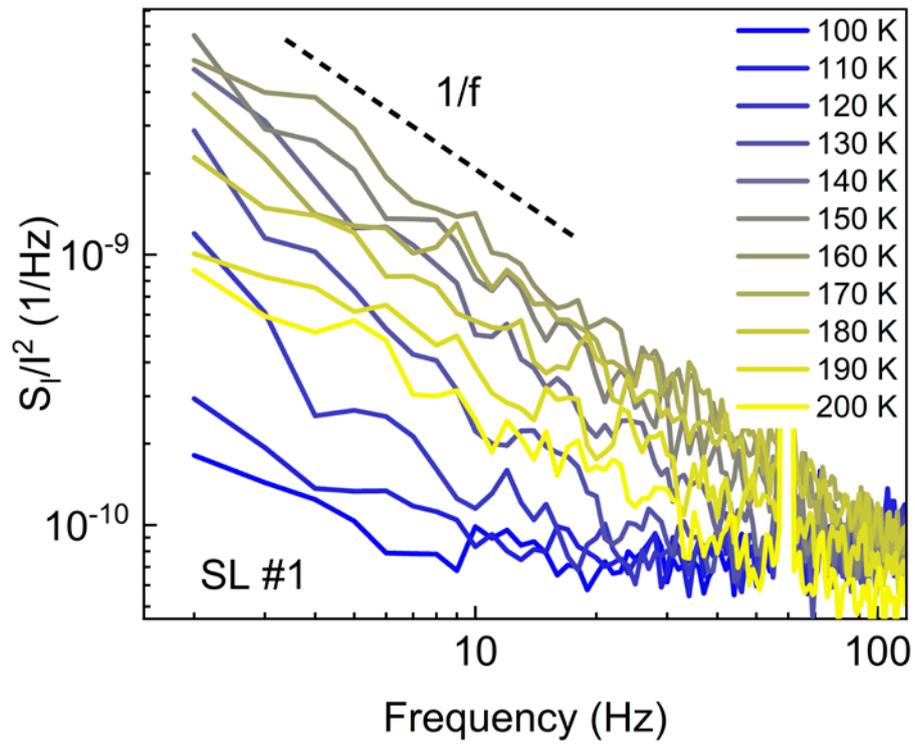

**Figure S4:** Normalized noise spectral density as a function of frequency measured at different temperatures for the ordered QD sample.

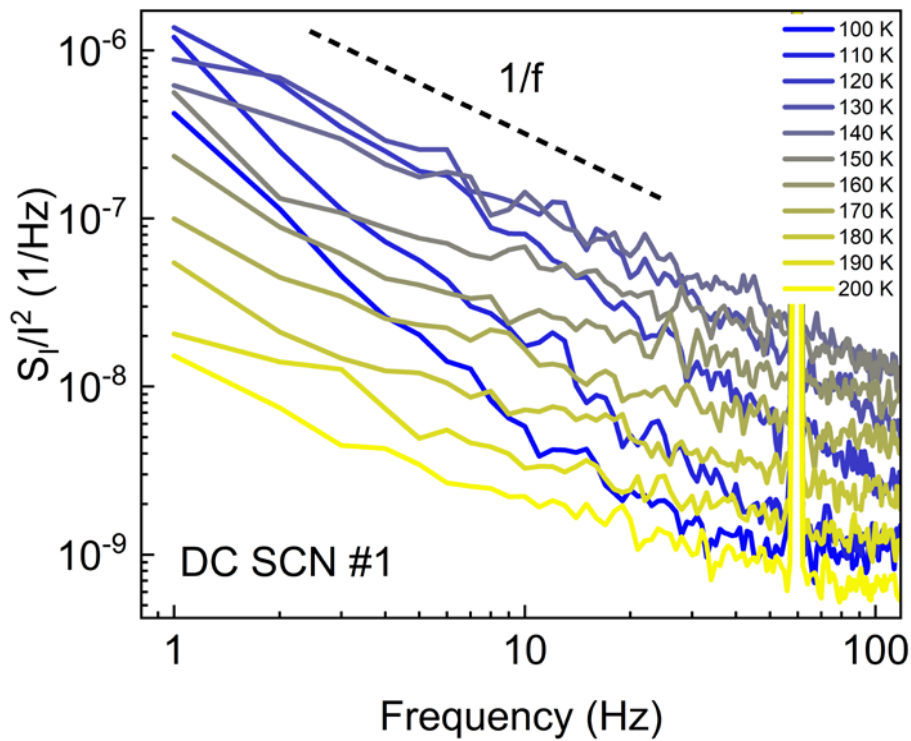

**Figure S5:** Normalized noise spectral density as a function of frequency measured at different temperatures for the random QD sample.